\documentclass[fleqn,usenatbib]{mnras}

\usepackage{newtxtext,newtxmath}
\usepackage{placeins}
\usepackage{comment}

\usepackage[T1]{fontenc}

\DeclareRobustCommand{\VAN}[3]{#2}
\let\VANthebibliography\thebibliography
\def\thebibliography{\DeclareRobustCommand{\VAN}[3]{##3}\VANthebibliography}

\usepackage{graphicx}	
\usepackage{amsmath}	
\usepackage{float}
\restylefloat{table}
\usepackage{tabularray}
\usepackage{caption}

\title[NEOWISE variability of black hole X-ray binaries]{Correlated mid-infrared and X-ray outbursts in black hole X-ray binaries: A new route to discovery in infrared surveys}

\author[C. John et al.]{
Chris John,$^{1}$
Kishalay De,$^{2,3}$\thanks{E-mail: kde1@mit.edu}
Matteo Lucchini,$^4$
Ehud Behar,$^{2,5}$
Erin Kara,$^2$
Morgan MacLeod,$^6$\newauthor
Christos Panagiotou,$^2$
and Jingyi Wang$^2$
\\
$^1$University of California, Berkeley, USA\\
$^{2}$MIT-Kavli Institute for Astrophysics and Space Research, 77 Massachusetts Ave., Cambridge, MA 02139, USA\\
$^{3}$NASA Einstein Fellow\\
$^{4}$Astronomical Institute, Anton Pannekoek, University of Amsterdam, Science Park 904, NL-1098 XH Amsterdam, Netherlands\\
$^5$Department of Physics, Technion, Haifa 32000, Israel\\
$^6$Center for Astrophysics | Harvard \& Smithsonian 60 Garden Street, MS-16, Cambridge, MA 02138, USA\\
}

\date{Accepted XXX. Received YYY; in original form ZZZ}

\pubyear{2015}

\begin{document}
\label{firstpage}
\pagerange{\pageref{firstpage}--\pageref{lastpage}}
\maketitle

\begin{abstract}
The mid-infrared (MIR; $\lambda \simeq 3 - 10\,\mu$m) bands offer a unique window into understanding accretion and its interplay with jet formation in Galactic black hole X-ray binaries (BHXRBs). Although extremely difficult to observe from the ground, the NEOWISE time domain survey offers an excellent data set to study MIR variability when combined with contemporaneous X-ray data from the MAXI all-sky survey over a $\approx 15$\,yr baseline. Using a new forced photometry pipeline that incorporates difference imaging for NEOWISE data, we present the first systematic study of BHXRB MIR variability in outburst. Analyzing a sample of 16 sources detected in NEOWISE, we show variability trends in the X-ray hardness and MIR spectral index wherein i) the MIR bands are typically dominated by jet emission during the hard states, constraining the electron power spectrum index to $p \approx 1 \text{--} 4$ in the optically thin regime and indicating emitting regions of a few tens of gravitational radii when evolving towards a flat spectrum, ii) the MIR luminosity ($L_{IR}$) scales as $L_{IR} \propto L_X^{0.82 \pm 0.12}$ with the $2-10$\,keV X-ray luminosity ($L_X$) in the hard state, consistent with its origin in a jet, and iii) the thermal disk emission dominates the soft state as the jet switches off and dramatically suppresses ($\gtrsim 10\times$) the MIR emission into a inverted spectrum ($\alpha \approx -1$, where $F_\nu \propto \nu^{-\alpha}$). We highlight a population of `mini' BHXRB outbursts detected in NEOWISE (including two previously unreported episodes in MAXI\,J1828-249) but missed in MAXI due to their faint fluxes or source confusion, exhibiting MIR spectral indices suggestive of thermal emission from a large outer disk. We highlight that upcoming IR surveys and the {\it Rubin} observatory will be powerful discovery engines for the distinctively large amplitude and long-lived outbursts of BHXRBs, as an independent discovery route to X-ray monitors. 
\end{abstract}




\section{Introduction}

Black hole (BH) X-ray binaries (XRBs) consist of a stellar mass BH accreting from a companion star, resulting in persistent or transient periods of luminous emission across the electromagnetic spectrum. The accretion process is most easily observed during outbursts, where an accretion disk can be formed around the BH, possibly with a collimated jet that carries away material from the central flow \citep{Fender2006}. Over the duration of an outburst, a BHXRB can exhibit dramatic changes in its accretion flow structure \citep{Homan2005}, correlated with geometry of the mass outflow in the form of winds or jets \citep{Ponti2016, Fender2016}. The physics of jet formation, and its connection to the accretion geometry near the central BH remains an active topic of research, but can directly probed with multi-wavelength observations spanning human-observable timescales \citep{Fender2004, Gilfanov2010, Russell2020, Trujillo2024}. The amount of mass and energy carried away in the jet provides an important benchmark for understanding the broader field of BH accretion, including its manifestations in super-massive BH systems that can impact entire Galactic ecosystems \citep{Kormendy2013, Heckman2014}. 

An outburst typically begins in the `hard' state, consisting of a geometrically thick, hot accretion flow together with a compact jet that is detectable from infrared (IR) to radio wavelengths \citep{Corbel2002, Tetarenko2015}. The jet emission characteristically exhibits a power law spectrum with a nearly flat synchrotron spectral index ($F_\nu \propto \nu^{-\alpha}$, where $\alpha \sim 0$ here) from low radio frequencies up to a characteristic break frequency ($\nu_b$) where the spectrum transitions to being optically thin with $\alpha \approx 0.6$ \citep{Fender2004}. As the outburst evolves and the accretion rate increases, the source transitions to the soft state where $\nu_b$ is observed to rapidly shift from the IR bands ($\nu \sim 10^{14}$\,Hz) to the radio bands ($\nu \sim 10^9$\,Hz), indicative of a rapid outward movement in the jet launching location \citep{vanderHorst2013, Russell2014, Zdziarski2022}. The compact jet and its associated emission is largely suppressed during the soft state, possibly due to the jet particle acceleration region moving too far away from the accretion disk \citep{Russell_2013, Romero2017}. Simultaneously, an optically thick, geometrically thin disk dominates the source spectrum during the soft state from IR to X-ray bands \citep{Shakura1973}. The sources transition back into the hard state towards the end of an outburst, where a compact jet is observed again \citep{Maccarone2003}.

During this typical evolution, the MIR bands (here defined as $\lambda \simeq 3-10\,\mu$m, corresponding to $\nu \simeq (0.3 - 1.0) \times 10^{14}$\,Hz) provide a unique window into the evolving emission components in the spectrum. Prior works have consistently detected excess mid-infrared emission in BHXRBs in the quiescent state (primarily using the {\it Spitzer} space telescope and the WISE all-sky survey; e.g. \citealt{Gallo2007, Gelino2010, Wang2014}), attributed to circumbinary dust disks \citep{Muno2006} or more likely to jet synchrotron emission \citep{Gallo2007}. During the outburst, the MIR bands land near the peak of the jet non-thermal spectrum (typically on the optically thin tail) during the hard state, while it becomes dominated by the disk emission during the soft state \citep{Russell2020, Trujillo2024}. Using cryogenic WISE mission data spanning $\approx 1$\,day, \citet{Gandhi2011} reported rapid (minutes to hours) MIR spectral slope changes in the BHXRB GX\,339-4 (1H\,1659-487) indicative of fast fluctuations in the jet acceleration region (see also \citealt{Baglio2018, Gandhi2017, Vincentelli2021}). Although extremely limited in the MIR, IR variability offers a unique perspective into jet formation in BHXRBs.


Previous studies of the IR variability BHXRBs have focused primarily on the near-IR bands that are observable from the ground ($\lambda \approx 1-2\,\mu$m; e.g. \citealt{Russell2006, Russell2007, Coriat2009, Russell2010}). However, this frequency range ($\nu \gtrsim 10^{14}$\,Hz) is consistently contaminated by both the disk and jet emission during the evolution of an outburst, rendering it non-trivial to separate the contribution of the two components (e.g. \citealt{Russell2006, Coriat2009}). Simultaneous monitoring campaigns reveal that the disk contribution is often negligible in the hard state at marginally longer wavelengths ($\nu < 10^{14}$\,Hz; $\lambda >3\,\mu$m, but difficult to observe from the ground), while being dominated by disk emission in the soft state when the jet switches off \citep{Russell2020, Trujillo2024}, allowing one to completely separate the jet and disk contributions to probe into jet formation physics. Finally, as optical and IR surveys become increasingly sensitive in the next decade, understanding correlated IR and X-ray activity offer new opportunities to discover BHXRBs independent of X-ray monitoring facilities \citep{Blackmon2023, Wang2024}. 

The WISE satellite \citep{Wright2010}, re-initiated as the NEOWISE mission \citep{Mainzer2014} has been performing an all-sky time domain survey in the $W1$ ($3.4\,\mu$m) and $W2$ ($4.6\,\mu$m) bands since 2014. The all-sky coverage and uniform cadence ($\approx 6$\,months) of NEOWISE makes it very well suited for time domain studies of slow Galactic variables and transients (e.g. \citealt{Zuckerman2023,Tran2024}), although the lack of forced photometry or difference imaging for dense Galactic plane fields has limited prior time domain applications. Owing to the unique NEOWISE MIR wavelength coverage that is well suited to study the evolution of jet emission in BHXRBs, we present the first systematic MIR study of jet evolution in BHXRBs using a custom photometry pipeline for NEOWISE data. We describe our data selection techniques in Section \ref{sec:data}, including historical X-ray and MIR light curves. In Section \ref{sec:analysis}, we analyze the correlated MIR and X-ray behavior of BHXRBs from the dataset, demonstrating the existence of correlated spectral changes between MIR and X-ray bands associated with the accretion flow. We discuss the implications for the observed evolution, and potential future applications in Section \ref{sec:disc}, and end with a summary in Section \ref{sec:summary}.

\section{Dataset}
\label{sec:data}
\subsection{NEOWISE MIR photometry}

The parent sample of sources for this study is taken from the catalog BHXRBs in the BlackCAT catalog\footnote{\url{https://www.astro.puc.cl/BlackCAT/transients.php}} \citep{Blackcat}. We include all sources with a reported optical/infrared counterpart during the outburst to include for forced photometry in NEOWISE data. We nominally adopt the reported distances of the sources in this catalog. Our pipeline for NEOWISE photometry operates in two modes. The first is by direct aperture photometry on the \texttt{unwise} processing of the WISE images \citep{Lang2014, Meisner2018} that retain the native spatial resolution of the instrument. Since some BHXRBs are located in extremely crowded regions of the Galactic plane, we also perform edifference imaging of NEOWISE data with respect to AllWISE images (from 2009-2011) and carry our forced photometry in the difference images (as in \citealt{De2023}), as we are primarily interested in the transient emission during outbursts. We do not use multiple observations of the same source within $\approx 1$\,d acquired during each sky visit (separated by $\approx 0.5$\,yr), and analyze the slower long-term evolution.

\begin{table}[]
    \centering
    \small
    \begin{tabular}{ccccc}
    \hline
    \hline
         Source Name & Distance & $N_H$ & $N_{WISE}$ & $t_{MAXI}$  \\
         & kpc & $10^{22}$\,cm$^{-2}$ & & days\\
         \hline
AT\,2019wey       & $1.0$      & 0.93  & 10 &  $7$           \\
GRS\,1716-249     & $2.4 $   & 0.84 & 4 &  $5$       \\
GRS\,1739-278     & $7.2$  & 2.79 & 2 & $1$      \\
GRS\,1915+105     &       $9.0$  & 6.00 & 21 &$3$              \\
GS\,1354-64       &     $ 25.0$    & 0.93 & 2  &  $1$           \\
MAXI\,J1348-630   &  $2.2$    & 0.74 & 6 & $3$           \\
MAXI\,J1820+070   & $2.9$  & 0.17 & 11 & $1$   \\
Swift\,J1357.2-0933* & $2.3$  & 0.03 & 5 &  $14$             \\
Swift\,J1753.5-0127   &  $6.0$ & 0.42 & 18 &  $7$  \\
1H\,1659-487* & 5.0 & 1.12 & 6 & 1\\
4U\,1543-475$^\dagger$ & 7.5 & 0.47& 4 & 1\\
MAXI\,J1535-571$^\dagger$ & 4.1 & 4.19 & 3 & 1\\
V4641\,Sgr & 6.2 & 0.33 & 8 & 1 \\
MAXI\,J0637-430 & 8.7 & 0.09 & 1 & 1 \\
IGR\,J17091-3624$^\dagger$ & 11.0 & 1.79 & 4 & 1\\
MAXI\,J1828-249$^\dagger$ & 8.0 & 0.32 & 4 & 5 \\

\hline
    \end{tabular}
    \caption{Summary of sources included in this work. For each source, we indicate the adopted distance, the column density along the line of sight ($N_H$), the number of WISE detections where the source was detected with signal-to-noise-ratio $>3$ and the time-bin used to create the MAXI light curves. $^\dagger$ indicates sources that are detected in NEOWISE data but do not have any simultaneous detections in MAXI (and hence not used for correlated behavior analysis), while * indicates sources which have at least one NEOWISE outburst where a coincident outburst was not detected in MAXI data.}
    \label{tab:sources}
\end{table}

\begin{figure*}
  \includegraphics[width=1\linewidth]{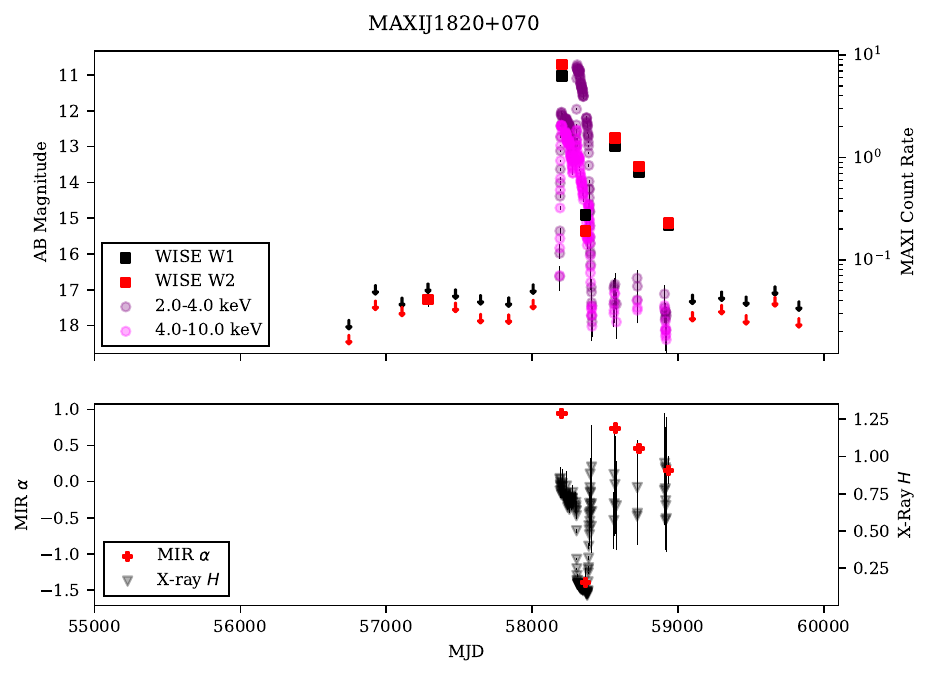}
  \caption{Combined MIR and X-ray light curve of the BHXRB MAXI\,J1820+070 from NEOWISE and MAXI. The top panel shows the multi-color MIR light curve (solid squares are detections and downward arrows are non-detections), as well as the X-ray light curve in the MAXI bands ($2-4$ and $4-10$\,keV; measured in MAXI count rates). We only show MAXI count rates during the epochs where the signal-to-noise ratio is greater than 3. The lower panel shows the measured spectral index ($\alpha$) in the MIR band and the X-ray hardness ratio ($H$) measured from the MAXI count rates.}
  \label{fig:maxij1820}
\end{figure*}

We visually examine all the light curves to search for known X-ray/optical/IR outbursts of BHXRBs and exclude any sources that are not detected (signal-to-noise-ratio $< 3$ at all epochs) or too heavily contaminated by nearby sources or artifacts. While some sources are nominally undetected/heavily contaminated using aperture photometry in dense Galactic fields, we use difference imaging photometry for these sources to retrieve useful data. As BHXRBs are typically too faint to be detected in quiescence at the WISE sensitivity threshold, the difference imaging photometry is reliable for our analysis of the transient accretion-related emission. Figure \ref{fig:maxij1820} shows an example of the NEOWISE light curve of the transient BHXRB MAXI J1820+070 from $\approx 13$\,years of observations. We also compute the spectral index $\alpha$ in the WISE bands for our analysis of the light curves. The resulting list of BHXRBs (16 sources) is given in Table \ref{tab:sources}, and the remaining light curves are shown in Appendix \ref{sec:simxrbs}.

\subsection{MAXI light curves}

With the objective of understanding MIR variability in BHXRBs correlated with the accretion disk variability, we retrieve X-ray light curves of source from the Monitor of All-sky X-ray Image (MAXI) mission \citep{Matsuoka2009}. The all-sky coverage, high sensitivity and temporal baseline that covers the entire NEOWISE mission duration makes it ideally suited for this analysis. We use the online MAXI on-demand light curve generator\footnote{\url{http://maxi.riken.jp/mxondem/}} service to generate X-ray light curves in two energy bands ($2-4$\,keV and $4-10$\,keV) for each source. We empirically adjust the time binning ($t_{MAXI}$; varied between 1 and 14 days) for each source to detect fainter outbursts for some sources, otherwise we nominally use a 1-day time bin. We use all measurements with signal-to-noise ratio $>3$, and define a X-ray hardness ($H$) as the ratio of the count rate in the $4-10$ to the $2-4$\,keV bands. The primary sample for understanding the correlated MIR and X-ray variability are objects which have at least one simultaneous NEOWISE and MAXI detection (12 sources); however, we also discuss a handful of sources that are detected in NEOWISE data but not in MAXI due to their faint X-ray flux and/or source confusion near the Galactic plane (Section \ref{sec:disc}).

\section{Analysis}
\label{sec:analysis}

\subsection{General trends in MIR and X-ray variability}
Figure \ref{fig:maxij1820} shows an example of a BHXRB light curve of MAXI\,J1820+070 detected simultaneously in NEOWISE and MAXI. As expected, outbursts are simultaneously detected in the MIR and X-ray bands, and consistently observed across the entire sample of sources barring constraints due to sensitivity. We use this source (Figure \ref{fig:maxij1820}) to demonstrate the characteristic correlated behavior; all the light curves are shown in Appendix \ref{sec:simxrbs}. While the source is undetected in quiescence, the MIR emission brightens by $> 6$\,mag at the start of the X-ray outburst (the rise time unresolved due to the slow cadence) followed by a slow decay. While both the X-ray count rate and MIR flux initially fade by $> 100\times$ within the first two WISE epochs ($\approx 6$\,months), we observe a resurgence in the MIR emission at $> 1$\,year coincident with the late-time decay and subsequent rebrightening episodes known in this source \citep{Ozbey2022}.

We attempt to understand this behavior by analyzing the correlated changes in the MIR spectral index and X-ray hardness during the same period. As shown in Figure \ref{fig:maxij1820}, the outburst begins with $H \approx 0.8$, indicating a fairly hard X-ray spectrum characteristic of the hard state seen in the early evolution of BHXRB outbursts (e.g. \citealt{Nakahira2010}), while $\alpha \approx 1.25$, indicating a steep\footnote{Throughout this paper, we refer to `steep' spectrum as a spectrum increasing in flux to lower frequencies and `inverted' spectrum as the opposite.} MIR spectrum. As the X-ray flux rises initially in the following months, we observe a dramatic increase in the $2-4$\,keV flux relative to $4-10$\,keV, causing the source to move to $H < 0.2$ when the source moves to the soft state. While the MIR flux drops by $> 4$\,mag during this phase, the spectral slope inverts to $\alpha\approx -1.5$, indicating a spectrum rising to higher frequencies. As the source subsequently fades in the X-ray, it transitions back to the hard state with $H \approx 0.75$ during its faint re-brightening episodes, and the total MIR flux rises again while the spectral slope becomes steep again with $\alpha \approx 1.0 - 1.2$. 

We interpret these trends noting that the MIR emission is expected to be dominated by the jet in the early outburst when the source is in the hard state. During this phase, the jet break frequency is generally at $\nu_b < 10^{14}$\,Hz (below the NEOWISE frequency bands; \citealt{Gandhi2011, Russell2020}) and therefore the MIR spectral slope traces the steep, optically thin part ($F_\nu \propto \nu^{-0.6}$) of the jet spectrum. As the source transitions into the soft state, the jet is expected to switch off, causing the MIR emission to drop precipitously as it becomes dominated by the Rayleigh-Jeans tail of the disk thermal emission ($F_\nu \propto \nu^{1/3}$ for a multi-color blackbody and $F_\nu \propto \nu^2$ for a pure blackbody; \citealt{Hynes2005, Pringle1981}). Therefore, we observe the transition to $\alpha \approx -1.2$ in this stage. Towards the end of the outburst and during its short re-brightening episodes, the X-ray source transitions back to the low, hard state \citep{Ozbey2022} causing the X-ray flux to remain low and the hardness to remain high, but the re-initiation of the jet emission causes the MIR flux to increase and exhibit a steep spectrum as the source subsequently fades into quiescence. 

\subsection{Correlated spectral slope changes}
\label{sec:corr_slope}
\begin{figure*}
    \centering
    \includegraphics[width=0.99\textwidth]{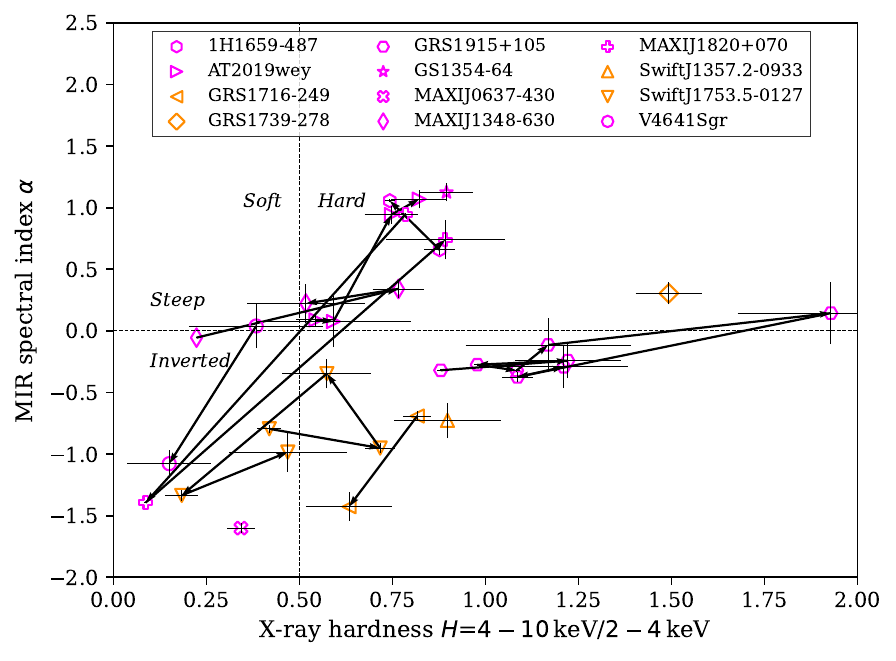}
    \caption{Correlated evolution of the MIR spectral index $\alpha$ with the X-ray hardness $H$ (based on the count rates in the two indicated energy bands) for the sample detected simultaneously in NEOWISE and MAXI. Symbols connected by arrows indicate the time evolution for each object (shown in the legend). The colors classify the objects into two distinctive behaviors -- with magenta symbols indicating objects where a transition in the X-ray state is accompanied by an inversion in $\alpha$, and the orange points indicating objects that do not show such drastic transitions (see text). The horizontal and vertical dashed lines show the adopted transition boundaries between steep/inverted MIR spectra and soft/hard X-ray spectra respectively.}
    \label{fig:mir_hardness}
\end{figure*}

Using the example of Figure \ref{fig:maxij1820}, we have discussed the characteristic behavior in correlated spectral and luminosity changes in BHXRBs over the duration of an outburst. The MIR spectral behavior can be understood by invoking an optically thin jet in the X-ray hard state, while emission from the disk dominates in the soft state producing an inverted spectrum. We now explore if this trend is universally seen across the sample of BHXRBs. In Figure \ref{fig:mir_hardness}, we show the evolution of the full sample of BHXRBs that are simultaneously detected in NEOWISE and MAXI, in the plane of X-ray hardness and MIR spectral index. For the majority of sources (9 out of 12 objects), we consistently observe state transitions that move diagonally in this plot, evolving from from steep to inverted MIR spectra as they move from the X-ray hard ($H \gtrsim 0.5$) to the X-ray soft ($H \lesssim 0.5$) state. Taking the MIR spectral slope to be on the optically thin part of the jet synchrotron spectrum when the MIR spectrum is steep, the MIR spectral index can be related to the power law index of the electron energy spectrum ($n \propto E^{-p}$; \citealt{RLbook})

\begin{equation}
    \alpha = \frac{p-1}{2}
\label{eq:alpha}
\end{equation}

From Figure \ref{fig:mir_hardness}, we observe the MIR spectral variability to be confined to $0 \lesssim \alpha \lesssim 1.5$ when $H > 0.5$, corresponding to $p \approx (1 - 4)$. We note that these measurements also assume that the MIR emission during the outburst is contributed only by the jet; therefore, the measurement would differ if the emission is significantly contaminated by a circumbinary disk \citep{Muno2006}, although they are expected to be relatively much fainter than the jet during the outburst \citep{Gallo2007, Gelino2010}. Although exhibiting a similar diagonal evolution in this phase space, a notable case is that of GRS\,1915+105 (Figure \ref{fig:GRS1915}), which transitioned from a high state to a low state in 2019, and is known for its remarkably hard color even in the soft, disk-dominated state (attributed to a high disk temperature; \citealt{Done2004}), producing the majority of points at $\alpha < 0$ and $H> 0.5$. Following its transition to the X-ray low state, it transitions into the $\alpha >0$ region indicative of a jet-dominated spectrum (consistent with \citealt{Imazato2021}).

A number of sources deviate from the characteristic behavior, including Swift\,J1357.2-0933 (Figure \ref{fig:ftoutbursts}), GRS\,1716-249 (Figure \ref{fig:grs1716-249}) and Swift\,J1753.5-0127 (Figure \ref{fig:swiftj1753}). These sources uniformly fall in the category of `failed transition' (FT) outbursts, where nearly the entire outburst was characterized by a low-hard or hard-intermediate state, without a transition into the X-ray soft state. Previous works for Swift\,J1753.5-0127 \citep{Shaw2019}, GRS\,1716-249 \citep{Saikia2022} and Swift\,J1357.2-0933 \citep{Russell2018} show that the disk remains a dominant component in the hard state even at MIR frequencies (producing an inverted spectrum). This is further supported by the repeated successful/failed outbursts of 1H\,1659-487 (GX\,339-4; Figure \ref{fig:1H1659}), where the two NEOWISE detections coincident with the hard states of the full outbursts (where the source does transition from the hard to the soft state) have steeper MIR spectral indices ($\alpha \approx 0.7 - 1.0$) compared to the NEOWISE observation during the failed outburst around MJD\,$58600$ (formally below our detection threshold and therefore not shown in Figure \ref{fig:mir_hardness}, but reported in the MAXI archive\footnote{\url{http://maxi.riken.jp/star_data/J1702-487/J1702-487.html}}) with $\alpha \approx 0.1$. Based on an analysis of FT outbursts from multiple sources, \citet{Alabarta2021} suggest that this is due to the mass distribution to be largely concentrated in the outer disk in FT outbursts, such that the optical/IR thermal emission is brighter and likely outshines the jet at these wavelengths.


\subsection{Global relationships between MIR and X-ray luminosity}

\begin{figure*}
  \includegraphics[width=1\linewidth]{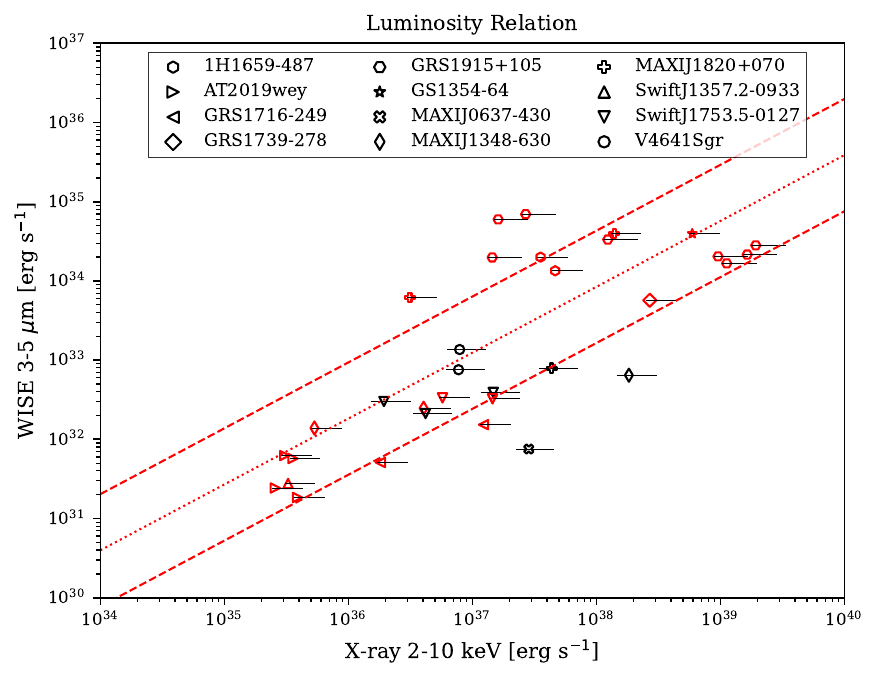}

  \caption{The dependence of MIR luminosity (in the WISE $3-5\,\mu$m band) on the X-ray luminosity (in the MAXI $2-10$\,keV band) for sources detected simultaneously in NEOWISE and MAXI, in the hard (red) and soft (black) states. The horizontal error bars indicate variations in the estimated X-ray luminosity for a range of spectral shape parameters (see text). The red dotted line indicates the best-fit power law for the hard state observations, while the dashed lines indicate the $1\sigma$ confidence region on the fit normalization. The best-fit hard state relationship is described by $L_{IR} [{\rm erg\,s}^{-1}] = 10^{2.98 \pm 0.71} L_X [{\rm erg\,s}^{-1}] ^{0.82 \pm 0.12}$, while the soft-state relationship is $L_{IR} [{\rm erg\,s}^{-1}] = 10^{28.58 \pm 0.38} L_X[{\rm erg\,s}^{-1}]^{0.11 \pm 0.28}$. 
  }
  \label{fig:HardXray-Relation} 
\end{figure*}


The simultaneous MIR and X-ray fluxes consistently indicate state-dependent luminosity changes due to the presence or absence of the jet. We quantify these trends by computing the MIR luminosity for each epoch where the sources are detected by performing a trapezoidal integration of the WISE band fluxes together with the assumed distance\footnote{We do not include interstellar dust extinction as it is expected to be insignificant at MIR wavelengths.} of the source (Table \ref{tab:sources}). For the X-ray data, we nominally assume that the source spectrum is a power law with with a photon index of $\Gamma = 1.7$ in the hard state, and a blackbody with $T = 1.0$\,keV in the soft state. We then use the \texttt{WebPIMMS} calculator with the previously estimated gas column density \citep{Blackcat} to convert the total MAXI $2.0 - 10.0$\,keV count rate into X-ray luminosity. We also include possible variations in $\Gamma$ (over $1.5 - 2.2$) and $T$ (over $0.5 - 2.0$\,keV) and adopt them as an uncertainty range in the X-ray luminosity. We caution that the sources certainly exhibit departures from these simple spectral models; however, since we aim to perform a homogeneous analysis of the MIR and X-ray luminosity trends (across a large dynamic range in X-ray fluxes), we nominally use this model in the absence of X-ray spectral data.

The evolution of the sample of BHXRBs in this plane is shown in Figure \ref{fig:HardXray-Relation}. We observe a strong correlation between the MIR and X-ray luminosity in the hard state (correlation coefficient $\approx 0.83$ between the logarithmic quantities), defined to be epochs where the X-ray hardness ($H$) is $> 0.5$, and is well-defined across four orders of magnitude in X-ray luminosity. Fitting a power-law to the trend, we find the relationship to be well described by $L_{IR} \propto L_X^{0.82 \pm 0.12}$ (shown in Figure \ref{fig:HardXray-Relation}). Comparing to the observed luminosity in the hard state, we find the MIR luminosity to be systematically suppressed $\sim 10\times$ at a fixed X-ray luminosity (Figure \ref{fig:HardXray-Relation}), consistent with the jet contribution switching off in the soft state while the source remains at high X-ray luminosity ($L_X \gtrsim 10^{37}$\,erg\,s$^{-1}$). The observed correlations are similar to those reported in previous works \citep{Russell2006, Russell2007, Homan2005}, and we discuss implications and potential applications in Section \ref{sec:disc}. However, we caution that due to the limited data availability, we only perform this analysis on the integrated luminosity in the WISE $3-5\,\mu$m and MAXI $2-10$\,keV bands; therefore the total MIR/X-ray luminosity may be underrepresented if a significant fraction falls outside the respective bands. Figure\,\ref{fig:HardXray-Relation} further shows that the observed MIR/X-ray luminosity ratio is typically $L_{IR}/L_X \sim 10^{-4}$, which is $\sim 10^3\times$ higher than the typical radio to X-ray luminosity ratio for BHXRBs ($L_{R}/L_X \sim 10^{-7}$ at a few GHz; e.g. \citealt{Koljonen2019}) in the hard jet-dominated state. This result is commensurate with the nearly flat spectrum, and demonstrates that although jets can be detected in the radio, the emitted power is dominated by the MIR and shorter wavelengths.

\section{Discussion}
\label{sec:disc}

\subsection{Constraints on particle acceleration from state-dependent MIR emission}

The correlated spectral transitions in MIR and X-ray emission during BHXRB outbursts conclusively support the scenario of MIR emission being dominated by jet activity. The observed spectral index in the MIR constrains the power-law index of synchrotron spectrum in the optically thin regime in most cases (where the spectrum is observed to be steep), and thereby the power-law index of the electron energy spectrum to be in the range $p \approx 1 - 3$. However, the particle distribution in the MIR-emitting regions of XRB jets are likely to be in the fast cooling regime \citep[e.g.][]{Pepe2015,Russell2020,Lucchini2021}, in which case the range of \textit{injected} particle slope would be in the range $p \approx 0 - 2$. Values of $p \approx 1 - 2$ favour a highly efficient acceleration mechanism such as reconnection, over shock acceleration \citep{Sironi2011,Sironi2014,Crumley2019}. On the other hand, the extreme cases of $p \approx 0$ observed in some sources in the hard state (see cluster of points near $\alpha = 0$ and $H > 0.5$ in Figure \ref{fig:mir_hardness}) is likely caused by synchrotron break frequency moving between or above the WISE bands, thus preventing an estimate of the spectral slope and particle spectrum index.  

The cases of the nearly flat MIR spectrum can be explained by the jet synchrotron emission being self-absorbed in this spectral band. An optically thick synchrotron source has a well defined surface brightness, which enables the estimate of the source size. The radius on the sky  $R_\mathrm{ssa}$ of a self absorbed synchrotron (disk shaped) source can be obtained from the synchrotron source function $S_\nu$ \citep[Eq.\,64 in][]{RLbook} and the luminosity density of the source $L_\nu$, as $R_\mathrm{ssa}=\sqrt{(L_\nu / S_\nu)} / 2\pi$, which can be expressed in km as

\begin{equation}
    R_\mathrm{ssa} = 940 \left( \frac{L_\mathrm{IR}}{10^{34}\mathrm{erg\,s}^{-1}}\right)^{1/2}
    \left( \frac{\lambda}{3\mu\mathrm{m}}\right)^{7/4}
    \left(\frac{B}{10^4\mathrm{G}} \right)^{1/4}
    \mathrm{km}
\label{eq:Rssa}
\end{equation}

\noindent 
where $L_\mathrm{IR}=\nu L_\nu$ is computed here at 
$10^{14}$\,Hz ($3\mu\mathrm{m}$). 
$R_\mathrm{ssa}$ can be seen to scale strongly with the observed wavelength $\lambda$, and weakly with the (poorly known) magnetic field $B$.
For the numerical pre-factor of Eq.\,(\ref{eq:Rssa})
we assumed a standard $p=2$ electron energy index (Eq.\,\ref{eq:alpha}).
The IR luminosity range (shown in Figure\,\ref{fig:HardXray-Relation}) indicate that the physical size of the IR emitting regions are merely of the order of Eq.\,(\ref{eq:Rssa}), namely compact jets of a few tens of gravitational radii. These calculations are consistent with the physical scale of the infrared jet forming region estimated by \citet{Chaty2011}.

Similarly, the global correlations in MIR and X-ray luminosity in the hard state can be compared to physical predictions for the scaling of the jet power with the mass accretion rate. \citet{Russell2006} and \citet{Heinz2003} show that the jet luminosity should scale as $L_{jet} \propto L_X^{0.7}$ in the radiatively inefficient regime for BHXRBs if the emission is dominated by a jet, while $L_{jet} \propto L_X ^{0.15 - 0.25}$ if the emission is dominated by a viscously heated disk. Our analysis therefore shows that the dependence of the MIR luminosity on the X-ray luminosity is consistent with the emission being dominated by a jet in the hard state ($L_{IR} \propto L_X^{0.82 \pm 0.12}$). We caution that the typically brief duration of the high state results in only a few NEOWISE observations coincident with the soft state. In addition, the limited high range of luminosity ($\gtrsim 10^{37}$\,erg\,s$^{-1}$) for the soft state results in nearly a flat relationship with the MIR luminosity. We note the specific cases of Swift\,J1753.5-0127 and V4641\,Sgr, which are known to exhibit exceptionally low luminosity soft states \citep{Shaw2016, Pahari2015} in this sample, and therefore appear as outliers to the generate soft state luminosity suppression (moving to the left and into the region occupied by the hard state relationship in Figure \ref{fig:HardXray-Relation}).

\subsection{Faint IR outbursts without MAXI detection}

\begin{figure*}
    \centering
    \includegraphics[width=\textwidth]{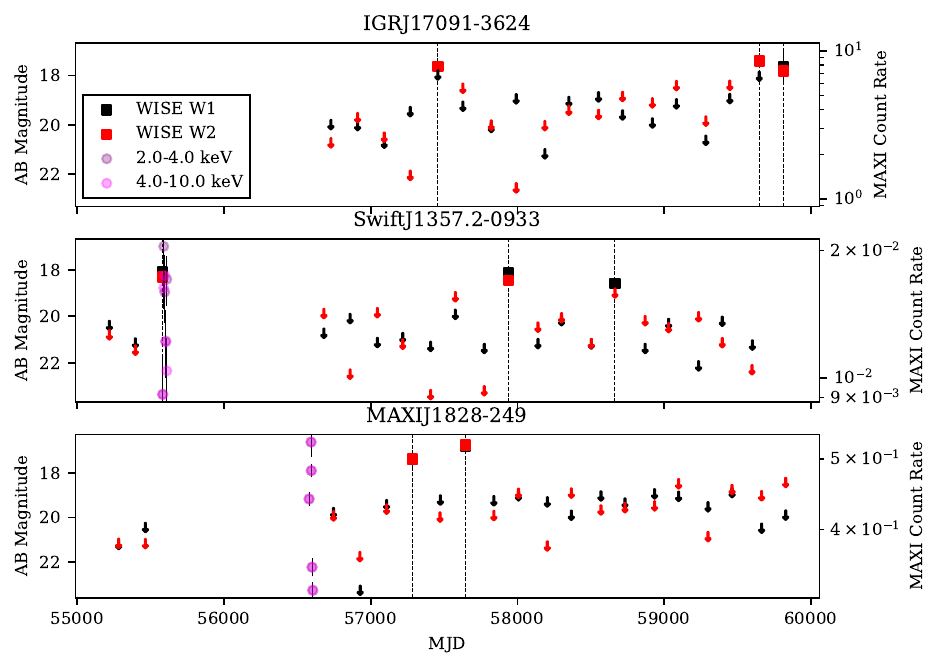}
    \caption{NEOWISE and MAXI light curves of BHXRBs with outbursts detected in the MIR but not simultaneously in the X-ray bands. In each panel, we show the WISE and MAXI data (as in Figure \ref{fig:maxij1820}) for the respective sources, with the WISE outbursts highlighted with black vertical dashed lines.}
    \label{fig:ftoutbursts}
\end{figure*}

While our analysis confirms the existence of correlated outbursts in IR and X-ray bands for BHXRBs that are simultaneously detected in NEOWISE and MAXI data, we discuss a sub-sample of outbursts where the faint MIR outbursts are clearly detected in NEOWISE data but are formally below the detection threshold of MAXI\footnote{While MAXI\,J1535-571 and 4U\,1543-475 do not have formally have simultaneous detection epochs in MAXI and NEOWISE, their light curves (Figures \ref{fig:maxij1535} and \ref{fig:4U1543}) clearly show that the MIR emission switches on at the tail end of the X-ray outbursts detected in MAXI when the sources transition to the low, hard state but below the MAXI detection threshold.}. In addition to the NEOWISE detection of the failed outburst of 1H\,1659-487 discussed in Section \ref{sec:corr_slope}, Figure \ref{fig:ftoutbursts} shows a collage of the light curves of additional similar outbursts as observed with NEOWISE and MAXI. Although not detected in MAXI data, some of these outbursts have been independently confirmed from other optical monitoring experiments. Notable examples include the multiple outbursts of Swift\,J1357.2-0933 in 2017 \citep{Drake2017} and 2019 \citep{vanVelzen2019}. We detect two previously unreported outbursts (Figure \ref{fig:ftoutbursts}) of MAXI\,J1828-249 (\citealt{Nakahira2013}; the discovery outburst was before the start of NEOWISE) in 2015 and 2016. The NEOWISE data show two outbursts of IGR\,J17091-3624 (Figure \ref{fig:ftoutbursts}) which are not detected in MAXI due to source confusion (from Sco\,X-2) but are previously reported from higher energy or optical bands \citep{Katoch2021, Saikia2022b}.



Previous works show that the known X-ray faint brightening episodes are associated with FT outbursts \citep{ArmasPadilla2014, Heinke2015} where the sources fail to transition into the soft state \citep{Alabarta2021}. The NEOWISE MIR colors indicate that nearly all the undetected X-ray outbursts are associated with flat/inverted spectra ($W1$ flux nearly equal to or brighter than $W2$). While the flat/inverted spectra can be explained by jet emission if the spectral peak lies $> 10^{14}$\,Hz, \citet{Alabarta2021} showed that the OIR emission is brighter for FT outbursts, and favour a larger outer disk where the mass is concentrated in the outskirts (and therefore prevents a full outburst) that can explain this behavior via thermal disk emission. Our analysis further shows that the faint `hard-only' re-brightening episodes seen in some objects (e.g. MAXI\,J1820+070; Figure \ref{fig:maxij1820}) immediately following a full outburst also produce luminous, steep spectrum MIR emission indicative of a dominant jet contribution, and are therefore different from FT outbursts where the jet likely remains sub-dominant to the outer thermal disk emission.

A related example is that of the BHXRB AT\,2019wey, which was first detected in the ATLAS optical survey (\citealt{Tonry2018}; also detected in NEOWISE before MAXI, see Figure \ref{fig:at2019wey}) several months before the eROSITA all-sky survey \citep{Predehl2021} reported a serendipitous X-ray detection, prompting multi-wavelength follow-up \citep{Yao_2021}. In NEOWISE data, AT\,2019wey transitions from being a flat spectrum source (indicating contamination from the disk; optical and radio analysis by \citet{Yao_2021} confirm that the break frequency was $\ll 10^{14}$\,Hz) to a steep spectrum source (likely dominated by the jet given the radio brightening reported by \citealt{Yao_2021}) as it transitions to a full bright X-ray outburst. These phenomena have recently motivated independent targeted optical searches \citep{Russell2019} to detect onset of outbursts prior to the rapid rise of the X-ray emission, as well as synoptic searches \citep{Wang2024} for new outbursts from known sources. However, they lack the temporal baseline and infrared sensitivity of NEOWISE.

\subsection{Discovering BHXRB outbursts with optical/IR surveys}

\begin{figure*}
    \centering
    \includegraphics[width=\textwidth]{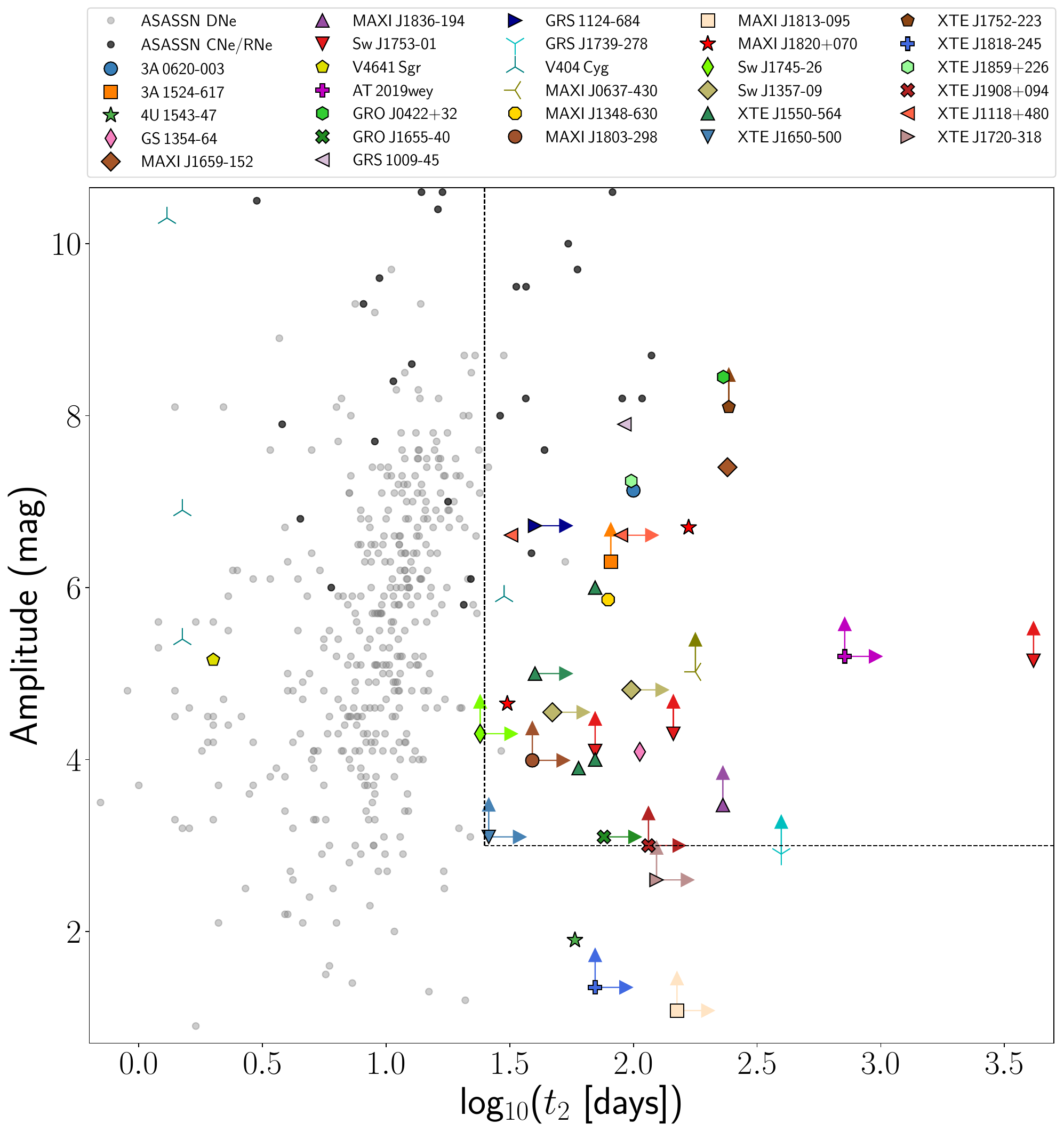}
    \caption{Phase space of optical/near-IR outburst amplitudes and durations (quantified by $t_2$, the time to decay by more than two magnitudes from peak) for known BHXRBs and dwarf/classical novae. The sample of dwarf (DNe) and classical/recurrent novae (CNe/RNe) are taken from the ASASSN survey compilation \citep{Kawash2021}. The outburst data for BHXRBs are taken from the \texttt{BlackCAT} compilation \citep{Blackcat} and references therein for 3A\,0620-003 \citep{Robertson1976}, 3A\,1524-617 \citep{Murdin1977}, 4U\,1543-47 \citep{Blissett1983}, GS\,1354-64 \citep{Kitamoto1990}, MAXI\,J1659-152 \citep{Russell2010b}, MAXI\,J1836-194 \citep{Russell2014b}, Swift\,J1753-0127 \citep{Halpern2005}, V4641\,Sgr \citep{Orosz2001}, AT\,2019wey \citep{Yao_2021}, GRO\,J0422+32 \citep{Castro-Tirado1992}, GRO\,J1655-40 \citep{Bailyn1995}, GRS\,1009-45 \citep{dellaValle1997}, GRS\,1124-684 \citep{dellaValle1991}, GRS\,J1739-278 \citep{Marti1997}, V404\,Cyg \citep{Casares1991}, MAXI\,J0637-430 \citep{Baglio2020}, MAXI\,J1348-630 \citep{Baglio2019}, MAXI\,J1803-298 \citep{Hosokawa2021}, MAXI\,J1813-095 \citep{ArmasPadilla2019}, MAXI\,J1820+07 \citep{Tucker2018}, Swift\,J1745-26 \citep{MunozDarias2013}, Swift\,J1357.2-0933 \citep{Rau2011}, XTE\,J1550-564 \citep{Orosz2002}, XTE\,J1650-500 \citep{Curran2012}, XTE\,J1752-223 \citep{Torres2009}, XTE\,J1818-245 \citep{Steeghs2005}, XTE\,J1859+226 \citep{Zurita2002}, XTE\,J1908+094 \citep{Chaty2006}, XTE\,J1118+480 \citep{Torres2002} and XTE\,J1720-318 \citep{Nagata2003}. In cases where the coverage is incomplete, we show estimated limits on the amplitude duration with arrows. The dashed lines delineate outbursts with large amplitudes ($> 3$\,mag) and long duration ($t_2 > 30$\,d) that can be used to separate BHXRBs from most DNe/CNe/RNe.}
    \label{fig:dne_xrt_ps}
\end{figure*}

It is remarkable that the NEOWISE dataset is enabling discoveries of new, previously unknown outbursts of BHXRBs given its very slow cadence and moderate sensitivity. This is primarily because BHXRB outbursts tend to be generally long-lived ($\gtrsim 50-100$\,d) and bright in the IR (where extinction is low), opening up the intriguing possibility for discovering new BHXRB outbursts in slow cadence IR surveys independent of X-ray monitors. Probing this population of mini-outbursts holds clues to a key missing ingredient in the demographics of BHXRBs. In the canonical evolution for low mass X-ray binaries, the companions evolve to shorter periods with decreasing accretion rate until a period minimum of $\approx 80$\,min \citep{PortegiesZwart1997,Yungelson2006}. \citet{Knevitt2014} argued that the observed paucity of short period BH binaries (compared to theoretical predictions; \citealt{Yungelson2006}) could be explained as a sustained transition to radiatively inefficient accretion outbursts below orbital period of $\approx 4$\,hours. \citet{Arur2018} suggested that the observed period distribution of BHXRBs is consistent with $\approx 600$ Galactic BH systems with periods between $3 - 10$\,hours, and 
a much larger population of $\approx 3000$\,systems between $2-3$\,hours (that may not produce bright X-ray outbursts; see also \citealt{Wang2024}). 

The prospects for discovering outbursts in optical/IR surveys have been previously discussed by \citet{Blackmon2023} by demonstrating a correlation between the peak luminosity of optical/IR outbursts and the BHXRB orbital period (i.e. upcoming optical/IR surveys are sensitive to low luminosity outbursts at short periods). Similarly, \citet{Wang2024} show that surveys in the red optical and near-IR bands may probe a larger population of Galactic BHXRB outbursts in the coming decade due to the substantially lower dust extinction, while the possibly higher duty cycle for FT outbursts may offer better constraints into the Galactic BHXRB population \citep{MacLeod2023}. However, the daunting challenge will be of separating these outbursts from the formidable foreground of other Galactic outbursts, notably dwarf and classical novae \citep{Starrfield2016, De2021}. We therefore explore the utility of the long outburst duration and large amplitudes of BHXRBs to separate them from these common interlopers. In Figure \ref{fig:dne_xrt_ps}, we present a compilation of the optical/near-IR outburst amplitudes (in mag) and durations (quantified by $t_2$, the time taken to decay by 2\,mag from outburst peak) of all known BHXRBs, compared to those of optically discovered dwarf and classical novae from the ASASSN survey \citep{Kawash2021}. 

Figure \ref{fig:dne_xrt_ps} clearly shows that BHXRB outbursts preferentially occupy the phase space of large amplitude and long duration outbursts in the Galactic plane. Notable exceptions include the very fast duration outbursts of the microquasars V404\,Cyg \citep{Casares1991} and V4641\,Sgr \citep{Orosz2001}, and the low amplitude outburst of 4U\,1543-47 \citep{Buxton2004}. We find that most BHXRB outbursts can be separated from interlopers by selecting for sources with amplitude $\gtrsim 3$\,mag and duration $t_2 > 30$\,d. While historically time domain surveys have frequently avoided the Galactic plane, ongoing surveys such as the Zwicky Transient Facility (ZTF; \citealt{Bellm2019}) and the upcoming Rubin observatory \citep{Ivezic2019} offer promising avenues to detect this population. In particular, the slow cadence and exquisite depth in redder optical bands for the Rubin observatory's Legacy Survey of Space and Time is extremely well suited to detect a large fraction of this population in the Galactic plane \citep{Wang2024}. Combining the wavelength-dependent photometric behavior (e.g. rapid color changes during accretion state transitions) discussed here, the optical/IR fluxes could be used to constrain the luminosity and distance to rule out impostors like young stars (that show very slow, or no color evolution; e.g. \citealt{Hillenbrand2018, Tran2024}). Finally, the emergence of deep, wide near-IR surveys of the Galactic plane such as Gattini-IR \citep{De2020}, WINTER \citep{Lourie2020} and PRIME \citep{Kondo2023}, and MIR surveys such as the NEO-Surveyor mission \citep{Mainzer2023}, will complement optical searches by detecting outbursts in the farthest and dustiest regions of the plane.

\section{Summary}
\label{sec:summary}

We present the first systematic study of MIR light curves of Galactic BHXRBs using data from the NEOWISE survey. Utilizing a difference imaging method to recover faint transient sources in dense Galactic plane fields together with X-ray light curves from the MAXI all-sky survey, we identify variability trends that corroborate previous suggestions and provide new insights for BHXRB jets and discovery:
\begin{enumerate}
    \item The MIR emission typically exhibits a flat or steep spectrum in the X-ray hard state, with a spectral index of $\alpha \approx 0 - 1.5$, consistent with synchrotron emission from a jet. The combination of the nearly flat spectra ($\alpha \approx 0$) and IR luminosity ($\sim 10^{34}$\,erg\,s$^{-1}$) observed in some sources indicate compact IR emitting regions of a few tens of gravitational radii. In the soft state, the MIR luminosity decreases and the spectrum becomes inverted, consistent with thermal emission from a disk. 
    \item The MIR luminosity is strongly correlated with the X-ray luminosity over four orders of magnitude in the hard state ($L_{IR} \propto L_X^{0.82 \pm 0.12}$), consistent with emission from a compact jet, while the emission is suppressed (by $\sim 10\times$) in the soft state for most sources as the jet switches off, and the MIR emission likely becomes dominated by disk emission, consistent with the MIR spectral evolution.
    \item A small fraction of sources that exhibit FT outbursts exhibit MIR emission contaminated by thermal disk emission even in the hard state, consistent with previous suggestions \citep{Alabarta2021} that a large outer disk contributes significantly in the optical/IR bands. Our analysis shows that the MIR emission associated with faint re-brightening episodes immediately following a full outburst in some sources also exhibit characteristics of a dominant jet contribution, and are hence different from FT outbursts where the jet likely remains sub-dominant to the outer disk in the MIR bands.
    \item We report detection of multiple IR outbursts in NEOWISE data that are not detected in the MAXI survey due to constraints from sensitivity and source confusion. The majority of these outbursts are associated with previously reported FT outbursts where the sources remain at low X-ray luminosity. We highlight the discovery of two previously unreported outbursts of MAXI\,J1828-249 in 2015 and 2016 that were missed by X-ray all-sky monitors.
    \item The serendipitous IR detection of BHXRB outbursts that are undetected by X-ray all-sky monitors bodes well for upcoming searches in optical/IR bands. We highlight that the large amplitudes ($> 3$\,mag) and long duration ($t_2 > 30$\,d) of BHXRB outbursts can be used to distinguish them from common interlopers in upcoming optical and IR wide-field surveys. 
\end{enumerate}

Although the NEOWISE time domain coverage provides tantalizing evidence for X-ray state-dependent MIR variability, suggestive of accretion/jet related emission mechanisms, future spectroscopy with {\it JWST} provides the firm opportunity to test these hypothesis via MIR spectroscopy to probe contributions from potential circumbinary disks \citep{Muno2006, Rahoui2010}. The increasing sensitivity of optical and IR surveys offer a promising opportunity to discover perhaps the most common types of FT outbursts from BHXRBs ($\gtrsim 40$\%; \citealt{Tetarenko2016, Alabarta2021, Lucchini2023}) from the most common type of accreting BH binaries (at short periods; \citealt{Knevitt2014}). These sources can be subsequently confirmed with follow-up up in the X-ray bands (e.g. for confirmation with {\it Swift} or {\it NICER}; \citealt{Gehrels2004, Gendreau2016}) or optical spectroscopy (to rule out the distinctive spectral features seen in nova contaminants; Figure \ref{fig:dne_xrt_ps}), as well as with serendipitous coverage from sensitive all-sky probes such as the recent {\it Einstein} probe mission \citep{Yuan2022}.

\section*{Acknowledgements}

We thank S. R. Kulkarni for valuable discussions. K. D. was supported by NASA through the NASA Hubble Fellowship grant \#HST-HF2-51477.001 awarded by the Space Telescope Science Institute, which is operated by the Association of Universities for Research in Astronomy, Inc., for NASA, under contract NAS5-26555.
E.B. is grateful for the hospitality of the MIT Kavli Institute for Astrophysics and Space Research where this work was carried out.

\section*{Data Availability}
 
All the NEOWISE light curves will be made available as Supplementary material. The MAXI light curves are publicly available online.



\bibliographystyle{mnras}
\bibliography{bibliography}





\appendix
\section{Simultaneous MIR and X-ray light curves of BHXRBs}
\label{sec:simxrbs}

\begin{figure*}
  \includegraphics[height=0.45\textheight]{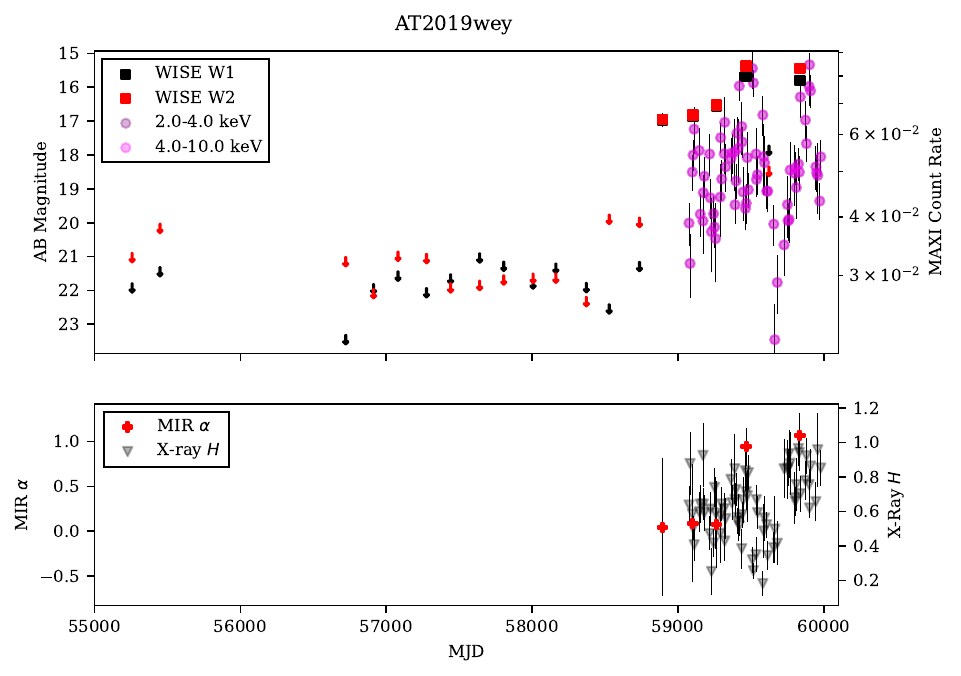}
  \caption{Combined MIR and X-ray light curve of the BHXRB AT\,2019wey from NEOWISE and MAXI, as in Figure \ref{fig:maxij1820}.}
  \label{fig:at2019wey}
\end{figure*}

\begin{figure*}
  \includegraphics[height=0.45\textheight]{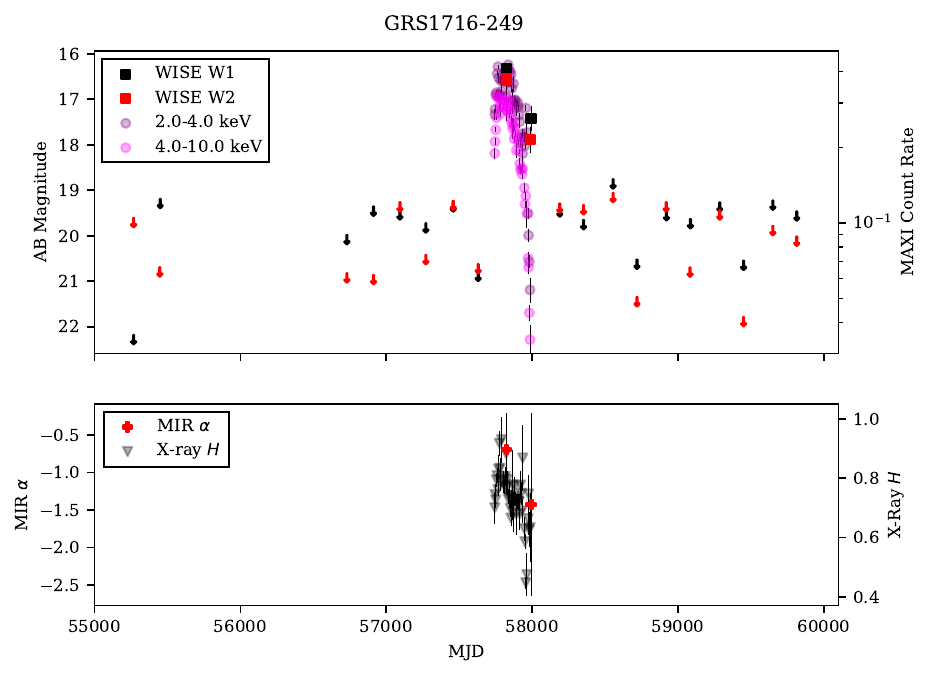}
  \caption{Combined MIR and X-ray light curve of the BHXRB GRS\,1716-249 from NEOWISE and MAXI, as in Figure \ref{fig:maxij1820}.}
  \label{fig:grs1716-249}
\end{figure*}

\begin{figure*}
  \includegraphics[height=0.45\textheight]{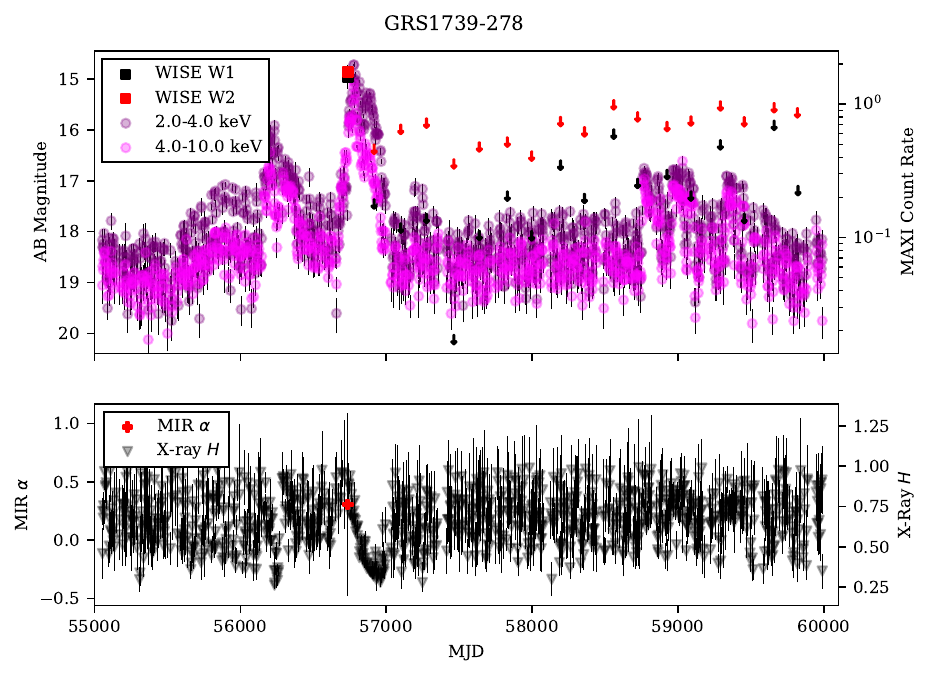}
  \caption{Combined MIR and X-ray light curve of the BHXRB GRS\,1739-278 from NEOWISE and MAXI, as in Figure \ref{fig:maxij1820}.}
  \label{fig:GRS1739-278}
\end{figure*}

\begin{figure*}
  \includegraphics[height=0.45\textheight]{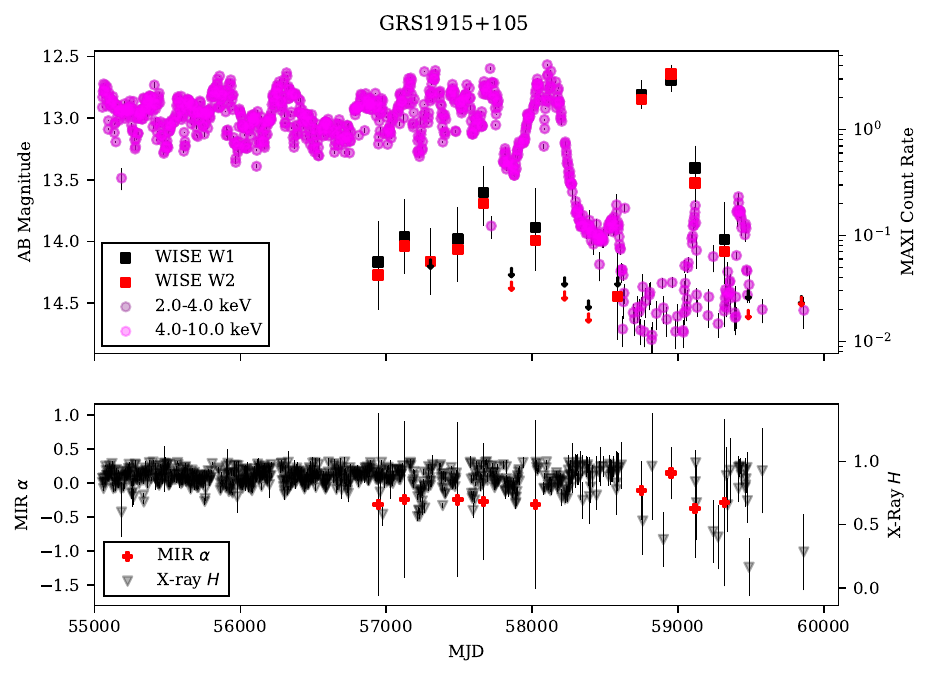}
  \caption{Combined MIR and X-ray light curve of the BHXRB GRS\,1915+105 from NEOWISE and MAXI, as in Figure \ref{fig:maxij1820}.}
  \label{fig:GRS1915}
\end{figure*}

\begin{figure*}
  \includegraphics[height=0.45\textheight]{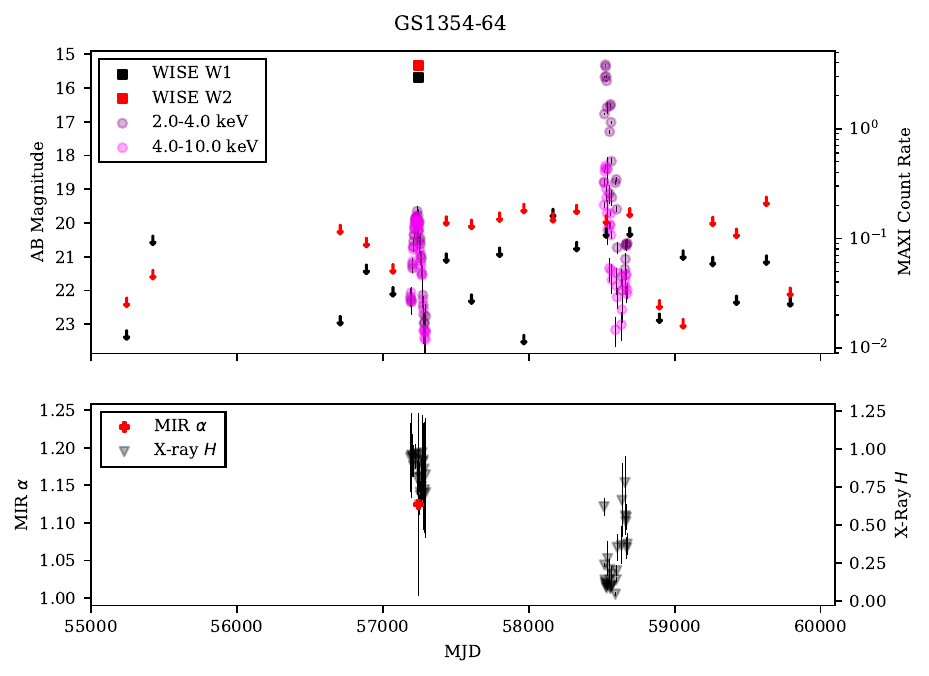}
  \caption{Combined MIR and X-ray light curve of the BHXRB GS\,1354-64 from NEOWISE and MAXI, as in Figure \ref{fig:maxij1820}.}
  \label{fig:GS1354}
\end{figure*}

\begin{figure*}
  \includegraphics[height=0.45\textheight]{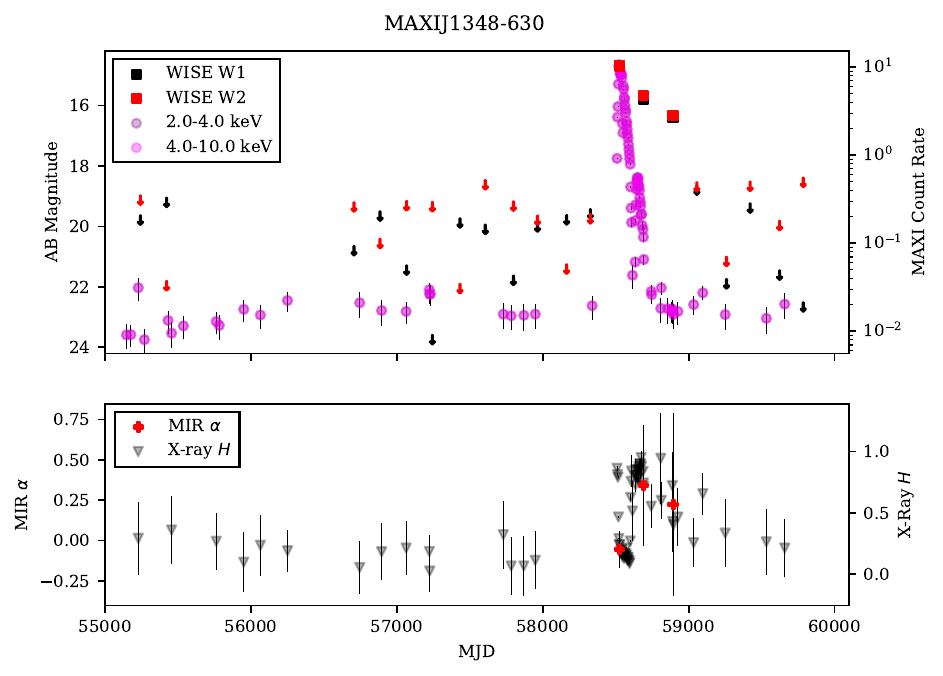}
  \caption{Combined MIR and X-ray light curve of the BHXRB MAXI\,J1348-630 from NEOWISE and MAXI, as in Figure \ref{fig:maxij1820}.}
  \label{fig:maxij1348}
\end{figure*}

\begin{figure*}
  \includegraphics[height=0.45\textheight]{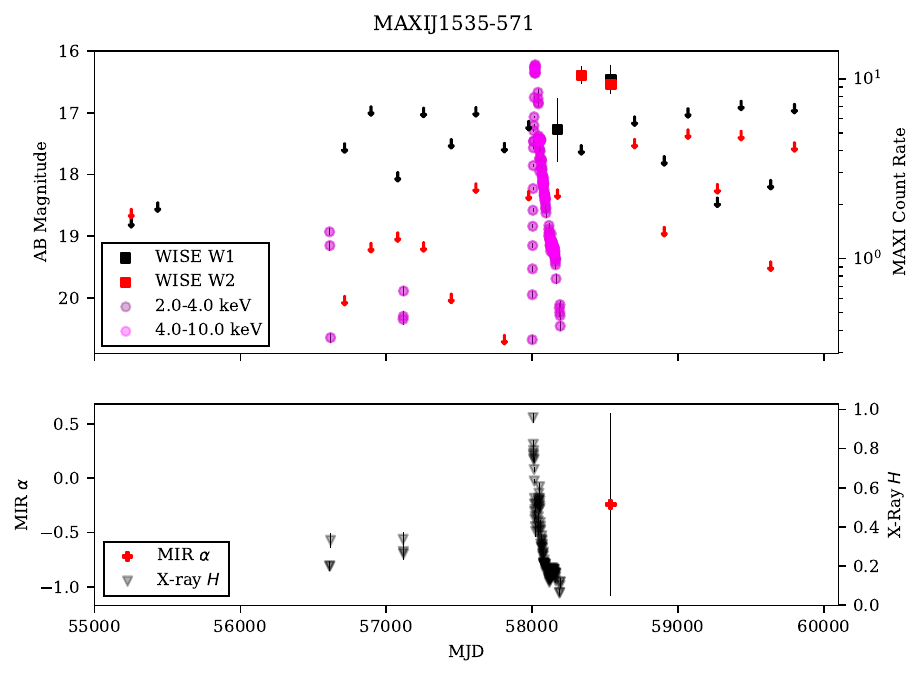}
  \caption{Combined MIR and X-ray light curve of the BHXRB MAXI\,J1535-571 from NEOWISE and MAXI, as in Figure \ref{fig:maxij1820}.}
  \label{fig:maxij1535}
\end{figure*}

\begin{figure*}
  \includegraphics[height=0.45\textheight]{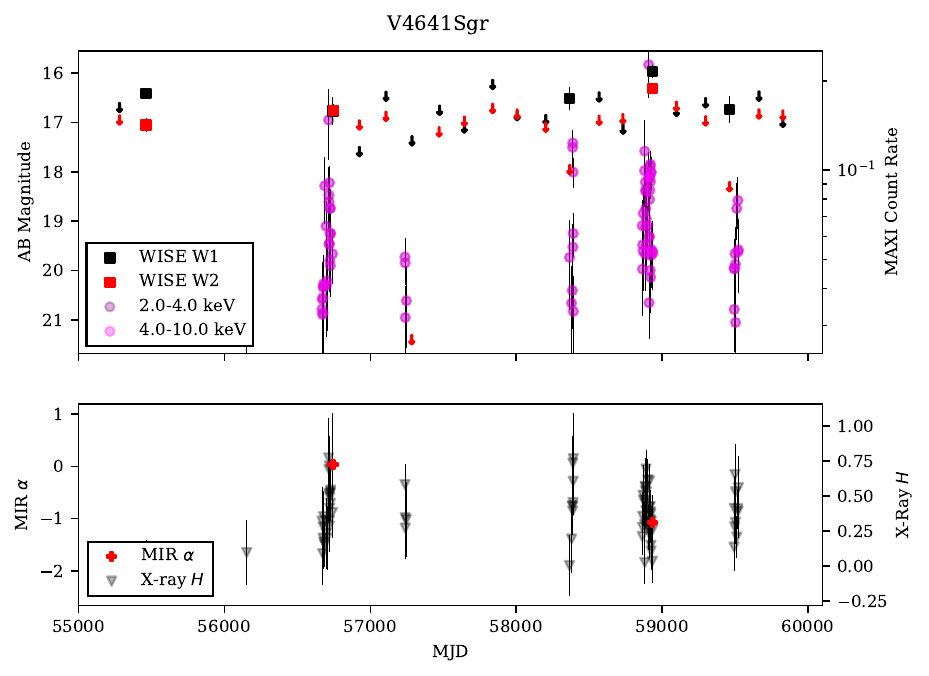}
  \caption{Combined MIR and X-ray light curve of the BHXRB V4641\,Sgr from NEOWISE and MAXI, as in Figure \ref{fig:maxij1820}.}
  \label{fig:V4641Sgr}
\end{figure*}

\begin{figure*}
  \includegraphics[height=0.45\textheight]{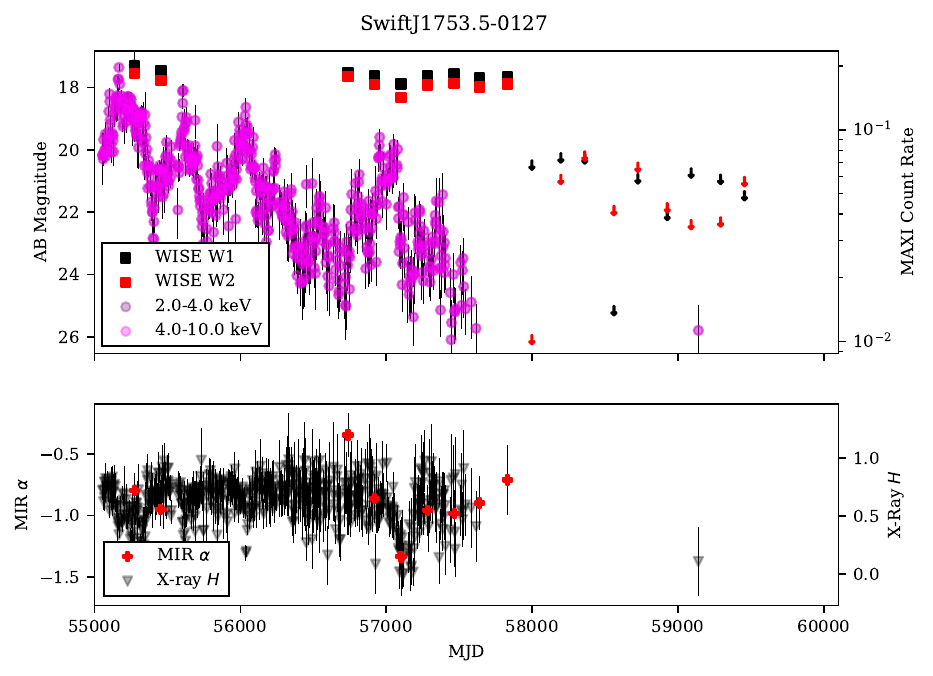}
  \caption{Combined MIR and X-ray light curve of the BHXRB Swift\,J1357.2-0933 from NEOWISE and MAXI, as in Figure \ref{fig:maxij1820}.}
  \label{fig:swiftj1753}
\end{figure*}

\begin{figure*}
  \includegraphics[height=0.45\textheight]{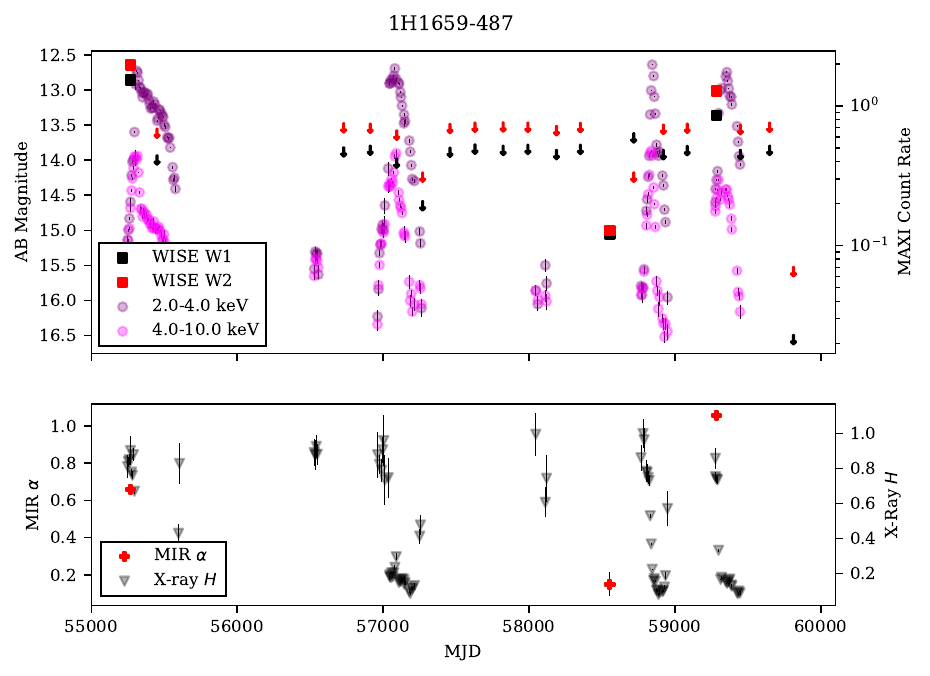}
  \caption{Combined MIR and X-ray light curve of the BHXRB 1H\,1659-487 from NEOWISE and MAXI, as in Figure \ref{fig:maxij1820}.}
  \label{fig:1H1659}
\end{figure*}

\begin{figure*}
  \includegraphics[height=0.45\textheight]{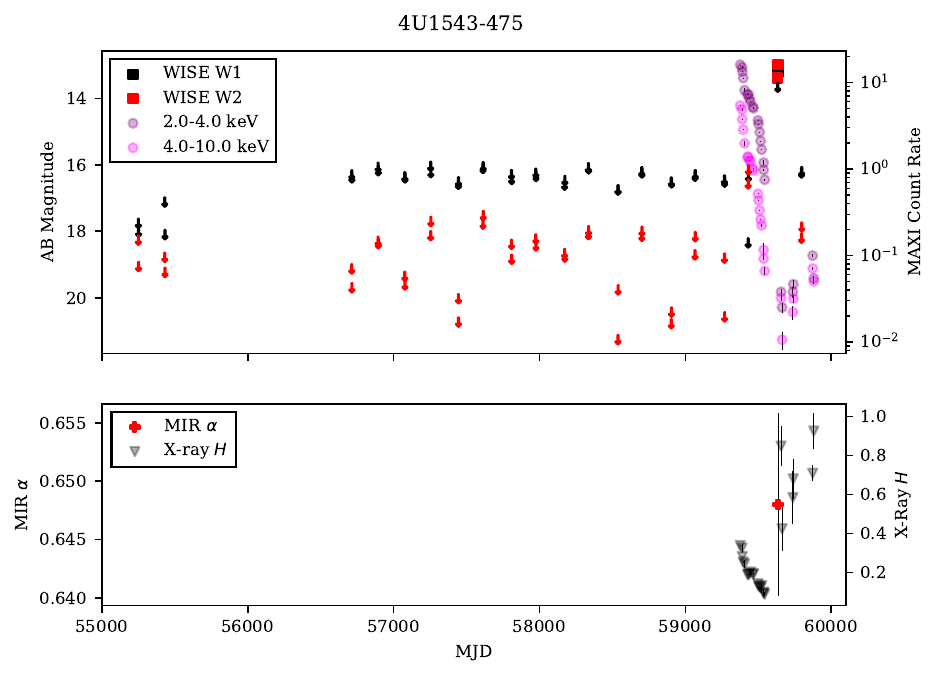}
  \caption{Combined MIR and X-ray light curve of the BHXRB 4U\,1543-475 from NEOWISE and MAXI, as in Figure \ref{fig:maxij1820}.}
  \label{fig:4U1543}
\end{figure*}

\begin{figure*}
  \includegraphics[height=0.45\textheight]{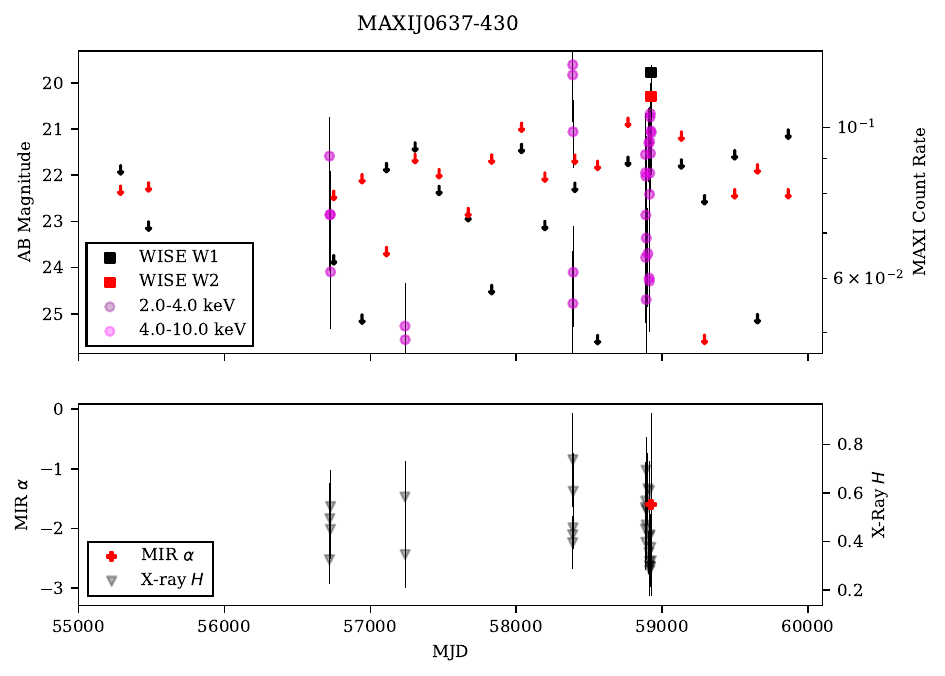}
  \caption{Combined MIR and X-ray light curve of the BHXRB MAXI\,J0637-430 from NEOWISE and MAXI, as in Figure \ref{fig:maxij1820}.}
  \label{fig:MAXIJ0637}
\end{figure*}

In Figures \ref{fig:at2019wey} to \ref{fig:MAXIJ0637}, we provide the list of combined NEOWISE and MAXI light curves for all sources used for analyzing the simultaneous MIR and X-ray behavior (i.e. sources that have at least one simultaneous NEOWISE and MAXI detection). 

\bsp	

\label{lastpage}

\end{document}